\documentstyle[epsfig,12pt]{article}
\def\thefootnote{\dag}
\newcommand{\be}{\begin{equation}}
\newcommand{\ee}{\end{equation}}
\newcommand{\bea}{\begin{eqnarray}}
\newcommand{\eea}{\end{eqnarray}}

\newcommand{\decays}{$B\to(K,K^*)(\ell^+\ell^-,\nu\bar\nu)$}

\begin{document}
\thispagestyle{empty}
~
\vspace{1truecm}

\hfill{
       \mbox{
             \begin{tabular}{l}
             {\bf Bari-TH/97-289}\\
             {\bf CPT-97/P.3560}\\
                  hep-ph/9711343
             \end{tabular}
}}

\vspace{1truecm}

\begin{center}
\begin{Large}
\begin{bf}
QCD Sum Rule Analysis of\\[.2cm] 
$B\to(K,K^*)(\ell^+\ell^-,\nu\bar\nu)$ \\[.2cm]
Decays\\[1cm]
\end{bf}
\end{Large}
\end{center}
\begin{center}
PIETRO SANTORELLI\\
\end{center}
\vspace{-.3truecm}
\begin{small}
\begin{it}
Dipartimento di Fisica, Universit\'a di Bari,
Istituto Nazionale di Fisica Nucleare, Sezione di Bari, Italy, and
Centre de Physique Th\'eorique-CNRS
%\footnote{On leave of absence from the University of Bari.}
, Marseille, France.\\
\end{it}
\end{small}

\vspace{1.5truecm}

\centerline{{\bf Abstract }}

\begin{small}
We use three-point QCD sum rules to calculate the form factors governing
the rare exclusive decays $B\to(K,K^*)\;\ell^+ \ell^-$,
$B\to(K,K^*)\;\nu \bar \nu$. We predict the branching ratios, the
invariant mass distributions of the lepton pair for $B\to(K,K^*)\;\ell^+
\ell^-$, and the spectra of missing energy for $B\to(K,K^*)\;\nu \bar
\nu$. The forward-backward asymmetry in $B \to K^* \ell^+ \ell^-$
provides us with interesting tests of the Standard Model and its
extensions.
\end{small}

\vspace{2truecm}

\begin{center}
Talk given at the\\
{\it IVth International Workshop on Progress in Heavy Quark Physics},\\
Rostock, Germany, September 20-22, 1997.\\
To appear in the Proceedings.
\end{center}

\newpage
\thispagestyle{empty}

~

\newpage

%\begin{center}
\begin{Large}
\begin{bf}
QCD Sum Rule Analysis of
%\\[.1cm] 
$B\to(K,K^*)(\ell^+\ell^-,\nu\bar\nu)$ \\[.1cm]
Decays\\[2mm]
\end{bf}
\end{Large}
%\end{center}
%
{\it Pietro Santorelli}\\[.5mm]
\begin{small}
Dipartimento di Fisica, Universit\'a di Bari,
Istituto Nazionale di Fisica Nucleare, Sezione di Bari, Italy, and
Centre de Physique Th\'eorique-CNRS, Marseille, France.\\
\end{small}
\vspace{-.3truecm}

{\bf Abstract }\\
{\small
We use three-point QCD sum rules to calculate the form factors governing
the rare exclusive decays $B\to(K,K^*)\;\ell^+ \ell^-$,
$B\to(K,K^*)\;\nu \bar \nu$. We predict the branching ratios, the
invariant mass distributions of the lepton pair for $B\to(K,K^*)\;\ell^+
\ell^-$, and the spectra of missing energy for $B\to(K,K^*)\;\nu \bar
\nu$. The forward-backward asymmetry in $B \to K^* \ell^+ \ell^-$
provides us with interesting tests of the Standard Model and its
extensions.
}

\vspace{-.5truecm}

\section*{Introduction}
\vspace{-.1truecm}
The rare $B$-meson decays induced by the flavour changing neutral
current $b \to s$ transition represent important channels for testing
the Standard Model (SM) and for searching for the effects of possible new
interactions \cite{ras}. In the SM these decays are forbidden at tree
level, and occur only through loop diagrams. For this reason, physics
beyond the SM can modify sensitively this kind of processes.\\
In order to test the SM predictions in the case of exclusive processes,
one needs to take into account nonperturbative QCD contributions
parameterized in terms of form factors. In this respect, the main
problem, in the case of \decays~decays, is the large kinematical range
for the squared momentum transfer ($q^2$) to the lepton pair, which
prevents us from making assumptions on the $q^2$ behaviour of hadronic
matrix elements. This difficulty can be overcome by using several
approaches, for example the three-point function QCD sum rule technique
\cite{shifman}, which is based on general features of QCD and allows us
to compute hadronic matrix elements in a large part of the $q^2$ range.
Three-point function QCD sum rules, first used to compute the pion form
factor \cite{ioffe}, have been applied to heavy meson semileptonic
\cite{neubertetal} and rare radiative decays \cite{coletal}. In the
following we discuss the results obtained applying this method to
calculate the relevant hadronic matrix elements in the \decays~decays.
\vspace{-.5truecm}
\section*{Effective Hamiltonian}
\vspace{-.1truecm}
The effective $ \Delta B =-1$, $\Delta S = 1$ Hamiltonian governing, in
the SM, the rare transition $b \to s \ell^+ \ell^-$ can be
written in terms of a set of local operators \cite{gri}: 
\begin{equation}
{\cal H}_W\,=\,4\,{G_F \over \sqrt{2}} V_{tb} V_{ts}^* \sum_{i=1}^{10} C_i(\mu)
O_i(\mu)\,,
\label{hamil} 
\end{equation}
\noindent 
where $G_F$ is the Fermi constant and $V_{ij}$ are elements of the
Cabibbo-Kobayashi-Maskawa matrix. We neglect terms proportional to
$V_{ub} V_{us}^*$ since the ratio $\left |V_{ub}V_{us}^*/
V_{tb}V_{ts}^*\right |$ is of the order $10^{-2}$. The operators $O_i$
are written in terms of quark and gluon fields and their expressions can
be found, for example, in Ref. \cite{BurasMunz}.  For the numerical
value of the Wilson coefficients $C_i(\mu)$ we follow the paper
\cite{BurasMunz}: the next-to-leading logarithmic corrections are
included only in the coefficient $C_9$, since at the leading
approximation $O_9$ is the only operator responsible of the transition
$b \rightarrow s~\ell^+ \ell^-$. Moreover, in our
numerical calculations we neglect the contribution coming from the
penguin operators, $O_3 \div O_6$.\\
The four-quark operators $O_1$ and $O_2$ generate both short- and  
long-distance (resonant) contributions to the processes. These
contributions can be taken into account by replacing $C_9$ with an
effective Wilson coefficient given by \cite{ali,odonnell}:
\begin{equation}
C_9^{eff}=C_9+(3 C_1+C_2)\left [h\left (\frac{m_c}{m_b},\frac{q^2}{m_b^2}
\right) 
+ { k} \; \sum_{i=1}^2
 {\pi \Gamma(\psi_i  \to
 \ell^+ \ell^-) M_{\psi_i} \over q^2-M_{\psi_i}^2+i M_{\psi_i} 
\Gamma_{\psi_i}} \right ] \,,
\label{c9eff} 
\end{equation}
with the parameter $k$ fixed according to the discussion in Ref. \cite{ali}. 
Finally, the effective $b\to s\ell^+\ell^-$  Hamiltonian can be recast in
the following form: 
\bea
{\cal H}_{eff}(b\to s\ell^+\ell^-)
  =  \frac{G_F}{\sqrt{2}}\frac{\alpha_{em}}{2\pi}V^*_{ts}V_{tb}
\left\{ \frac{}{} \right.& - & \left.\frac{2im_b}{q^2}
C_7(m_b)\left[\bar s\sigma_{\mu\nu}q^{\nu}(1+\gamma_5)b\right]
\left[\bar\ell\gamma^{\mu}\ell\right] \right. 
\nonumber \\ 
&+& C_9^{eff}(m_b) 
\left[\bar s\gamma_{\mu}(1-\gamma_5)b\right]
\left[\bar\ell\gamma^{\mu}\ell\right]
\nonumber\\
&+& \left . C_{10}(m_b) 
\left[\bar s\gamma_{\mu}(1-\gamma_5)b\right]
\left[\bar\ell\gamma^{\mu}\gamma_5\ell\right]\right \}\;.
\label{e:hamil1}
\eea
On the other hand, the $ b\to s \nu\bar\nu$ process is governed by the 
effective Hamiltonian
\be
{\cal H}_{eff}(b\to s \nu\bar\nu) = 
{G_F \over \sqrt 2} {\alpha_{em} \over 2 \pi \sin^2(\theta_W)}
\; V_{ts} V^*_{tb} \; X \!\left(\frac{m_t^2}{M_W^2}\right) \; 
\left[{\bar b} \gamma^\mu ( 1- \gamma_5) s \right]\; 
\left[{\bar \nu} \gamma_\mu ( 1- \gamma_5) \nu \right]
\label{e:hamil2}
\ee
obtained from $Z^0$ penguin and box diagrams where the dominant
contribution corresponds to a top  quark intermediate state ($\theta_W$
is the Weinberg angle). The leading and the ${\cal O}(\alpha_s)$
corrections, deriving from two-loop diagrams, are taken into account in
the $X_0$ and the $X_1$ term, respectively, of the function $X$:
\be
X(x)= X_0(x) + {\alpha_s \over 4 \pi} X_1(x) \;\;,
\ee
which can be found in \cite{inamibuchalla}.
\begin{table}
\begin{center}
\begin{small}
\begin{tabular}{|c|c|c||c|c|c|}
\hline \hline
   & $F(0)$ & $M_P \; (GeV)$ &    & $F(0)$ & $\beta \; (GeV^{-2})$  \\ 
\hline \hline
$F_1$ & $0.25 \pm 0.03$ & $5$ & 
$A_1$ & $0.37 \pm 0.03$ & $-0.023$ \\
\hline
$F_0$ & $0.25 \pm 0.03$ & $7$ &
$A_2$ & $0.40 \pm 0.03$ & $0.034$ \\
\hline
$V$ & $0.47 \pm 0.03$ & $5$ &
$T_2$ & $0.19 \pm 0.03$ & $-0.02$ \\
\hline
$A_0$ & $0.30 \pm 0.03$ & $4.8$ & 
$F_T$ & $-0.14 \pm 0.03$ &     \\
\hline
$T_1$ & $0.19 \pm 0.03$ &$5.3$ & 
 & &  \\
\hline
\end{tabular}
\end{small}
\end{center}
\vspace{-.5truecm}
\caption{{\small Parameters of the form factors. The functional $q^2$
dependence is either polar: $F(q^2)=F(0)/(1-q^2/M_P^2)$ or linear: 
$F(q^2)=F(0)(1+ \beta q^2)$.
}}
\end{table}

\vspace{-.5truecm}

\section*{$B\to(K,K^*)$ form factors}
\vspace{-.1truecm}
We have evaluated in Refs. \cite{noi1,A0} the matrix elements of the hadronic
operators appearing in Eqs.(\ref{e:hamil1}),(\ref{e:hamil2}) between B
and K, $K^*$ states, using three-point function sum rules. 
The definition of the various form factors can be found in \cite{noi1}.
The parameters  of the form factors are collected in Table 1. In Figure 1
we have plotted the $q^2$ behaviour of the form factors in the range
$[0,15]~GeV^2$. The form factor $T_3(q^2)$
can be obtained using the quark equation of motion
\def\thefootnote{\diamond}
\hspace{-.3truecm}
\footnote{However, the cancellation between the two terms in the
Eq.~(\ref{t3}) implies a large error in its determination. In any case,
if we limit to consider light leptons (i.e. excluding the case of $\tau$)
in $B\to (K,K^*)\ell^+\ell^-$ the contribution of $T_3$ is negligible. A
direct calculation of $T_3$ is in progress.}
\be
T_3(q^2)=-M_{K^*} (m_b-m_s) {A_3(q^2)-A_0(q^2) \over q^2}\;.
\label{t3} 
\ee
As discussed in Ref. \cite{noi1}, the same procedure can be successfully 
adopted for $F_T$, where the equation
\be
F_T(q^2) = \left(M_B+M_K\right)(m_b+m_s)\frac{F_0(q^2)-F_1(q^2)}{q^2}\,,
\ee 
gives results in agreement with the direct calculation, but with
a sensibly larger error.

Our results agree at $q^2=0$ with similar calculations performed using
the Light Cone Sum Rule (LCSR) approach \cite{Aliev}. It should be
stressed, however, that the $q^2$ behaviour predicted for the form
factors by LCSR is quite different, in particular for $A_1$ and $T_2$
which LCSR predict to be increasing functions of $q^2$, whereas in
three-point QCD sum rules they are quite flat. Various possible sources
for these differences have been advocated in the literature
\cite{braun97}; a discussion is beyond the aims of the present paper.
\begin{figure}
\centerline{\epsfig{file=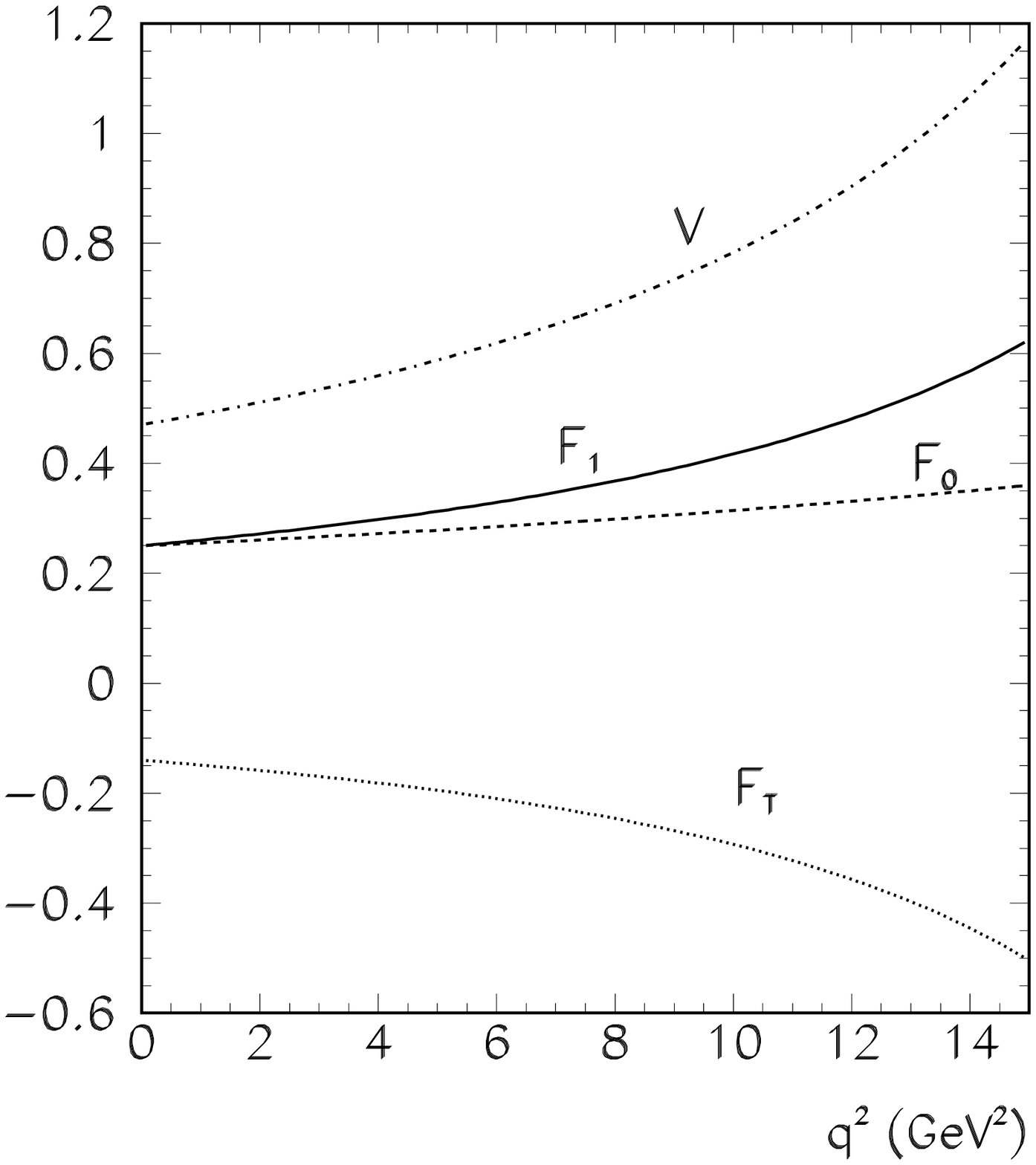,height=7truecm}
\epsfig{file=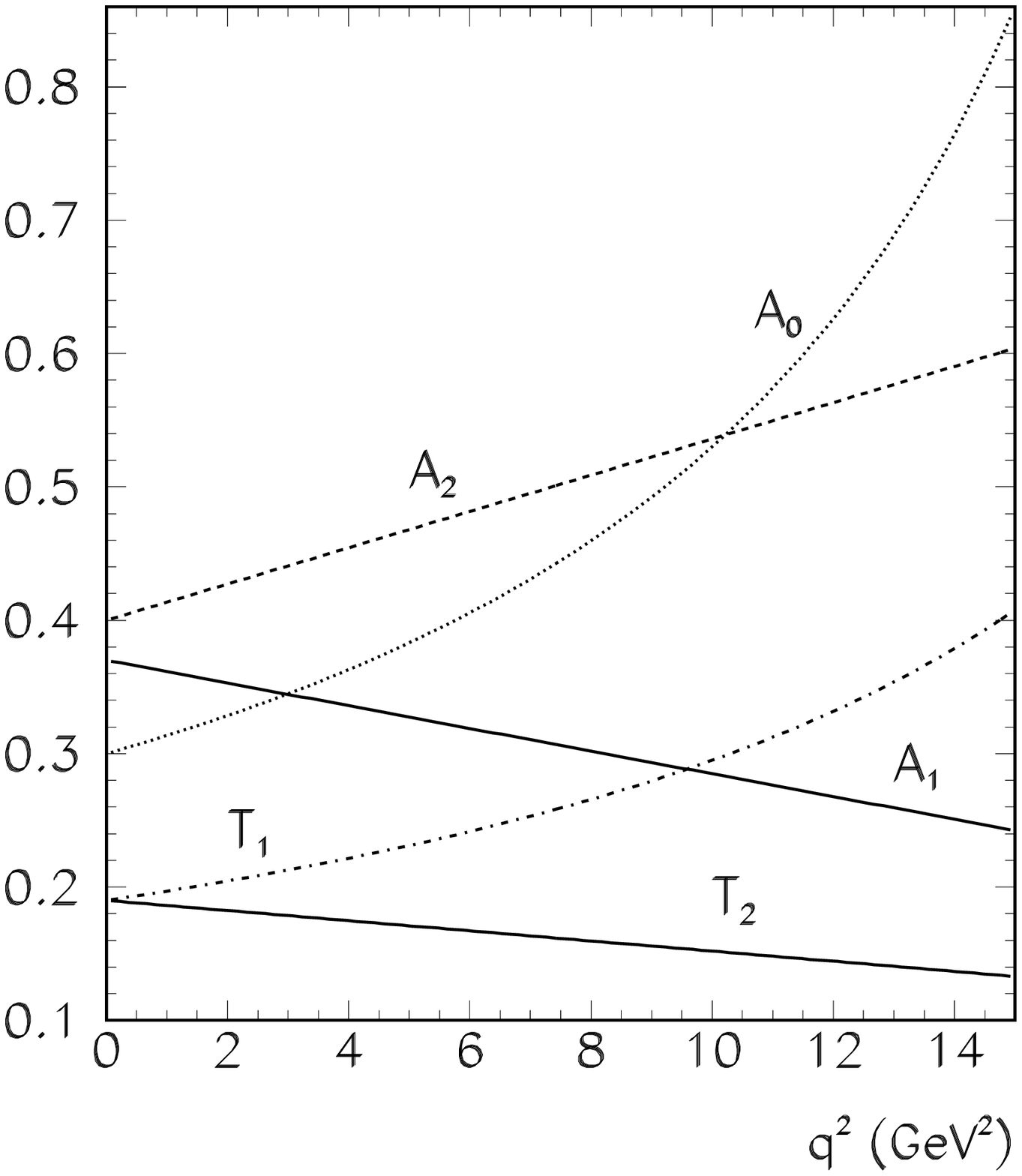,height=7truecm}}
\vspace{-.5truecm}
\caption{{\small $B\to(K,K^*)$ form factors. The curves refer to 
the parameters in Table 1.}} 
\end{figure}
\vspace{-.5truecm}
\section*{Predictions for \decays~processes}
\vspace{-.1truecm}
Using the computed $B\to(K,K^*)$ form factors, it is straightforward to
determine the differential decay rates and the asymmetries for the
\decays~processes. The various analytical expressions, together with the
numerical results, can be found in Refs.~\cite{noi1,noi2}. Here we only
report our predictions for the branching ratios, without including, for
the transition with $\ell^+\ell^-$ in the final state, the resonant
long-distance contributions, which are important only in narrow $q^2$
regions centered at $q^2=M^2_{ J/\psi,\psi^{\prime} }$. Using
$|V_{ts}|=0.04$, we predict the branching ratios reported in the
following table.
$$
\begin{small}
\begin{tabular}{|c|c|}
\hline 
Br($B\to K \ell^+\ell^-$) = 3 $\times~10^{-7}$ & Br($B\to K \sum_i
\bar\nu_i\nu_i$)  = $(2.4\;\pm\;0.6)~\times~10^{-6}$ \\  \hline 
Br($B\to K^*\ell^+\ell^-$) = 1 $\times~10^{-6}$ & Br($B\to K^*\sum_i
\bar\nu_i\nu_i$)  = $(5.1\;\pm\;0.8)~\times~10^{-6}$  \\
\hline 
\end{tabular}
\end{small}
$$
Branching ratios of this order of magnitude should be measured at the
future B factories, where $10^9~B\bar B$ pairs {\it per} year are
expected to be produced. In particular, the processes with neutrinos in
the final state are theoretically interesting, due to the absence of
long-distance resonant contributions and due to the fact that they are
induced, in SM, by only one operator; therefore, they represent good
candidates for testing the SM predictions and for probing the effects of
possible new interactions. Another important quantity is represented by
the forward-backward asymmetry in $B\to K^*\ell^+\ell^-$ transition,
which  plays a fundamental role in testing the Standard Model
predictions. It is expected that such an asymmetry should be about
$-0.60\%$ for large $q^2$, and therefore it should be measured at B
factories. Its importance come from the fact that it depends not only on
the magnitudes but also on the relative signs of the Wilson
coefficients, allowing to test their values as they are predicted by SM.
So any observed deviation from the predictions, assuming a theoretical
error of about 30\%, would be interpreted as a signal of New Physics.
\vspace{-.5truecm}
\section*{{\small {\bf Acknowledgments}}}
\vspace{-.1truecm}
I thank P. Colangelo, F. De Fazio and E. Scrimieri for their
collaboration on the topics discussed here.

\end{document}